\documentclass[showpacs,aps,prb]{revtex4}
 \usepackage{amsmath,amssymb,pstricks,pst-plot}

 \newcommand{\nts}[1]{\tmspace{-}{#1\thinmuskip}{#1\txtmu}}
\allowdisplaybreaks
\begin{document}
\title{Ground state energy of an interacting electron system in the background of two opposite magnetic strings}
% \title[Spectral properties of a two string  electron system]{Spectral properties of an electron system in the background of two opposite magnetic strings} 

% \author[J\"urged  Dietel]{J\"urgen  Dietel}
\author{J\"urgen  Dietel}
\affiliation{
Institut f\"ur Theoretische Physik, \\
 Freie Universit\"at Berlin, \\
 Arnimallee 14, D 14195 Berlin, Germany}
\email{dietel@physik.fu-berlin.de}

\begin{abstract}
Motivated by our earlier work, we show in this paper rigorously 
that the ground state energy and degeneracy 
of an infinitely extended system of interacting electrons in the background 
of a homogeneous magnetic field and two separated magnetic strings
of opposite strength
is the same as for the system without strings. 
By using symmetry considerations we obtain further that the energy spectrum 
does not depend on the string separation distance for strictly positive 
distances. As a side effect of our considerations, we obtain a 
virial theorem for the two string system in the case of a homogeneous 
interaction potential which has the same form as the virial theorem without 
the strings.  
\end{abstract}


\maketitle

\section{introduction}

Since the discovery of the Aharonov-Bohm \cite{ah1} effect there is a growing 
interest in quantum mechanical systems in the presence of     
a finite number of magnetic strings. 
Due to the localization of the magnetic field of a string these systems are 
rather quantum mechanical objects than classical ones. The magnetic strings 
affect the energy spectrum and the electronic wavefunctions  
due to the extension of the wavefunctions which is governed by 
the uncertainty principle.  
In this paper we consider an infinitely extended system  
of two dimensional interacting electrons under 
the influence of a homogeneous magnetic field $ B $ and two magnetic 
strings of opposite strength separated by a distance $ d $.  
We will show that the spectrum of the electron system 
does not depend on $ d $ for $ d >0 $. Furthermore, 
we will show that the ground state energy and the degeneracy of the electron 
system are the same with and without the two strings \cite{note}.  

We have two motivations to consider this two string system.
The first motivation comes from the observation \cite{hlr}, \cite{di1} 
that the Green's function 
of Chern-Simons theories for fractional quantum Hall systems in the 
commonly used Coulomb gauge vanishes
exponentially with an exponent proportional to the 
logarithm of the area of the system. 
We have analyzed in Ref. \onlinecite{di2} the physically 
better motivated Chern-Simons 
Green's function in the temporal gauge. The Green's function is 
divergent to lowest order 
perturbation theory at zero temperature.  
In contrast to this we have shown in Ref. \onlinecite{di2} 
non-perturbatively that in the case of the validity of the 
two string spectral properties mentioned above
we obtain a finite Green's function in the temporal gauge.
It was shown that for non-interacting electrons 
the ground state energy is the same with and without the two strings. 
Furthermore, we calculated  by an explicit construction 
of the ground state wavefunctions the interaction free 
Chern-Simons Green's function in the temporal gauge.
At last, we gave two non-rigorous physical arguments  
that the interacting electron system should also have the same 
ground state energy with and without the two strings.
This will be shown rigorously in this paper.  

The second motivation for this paper comes from the observation made in 
Ref. \onlinecite{ka1} that in the continuum limit of a tight 
binding model, the force   
of a screw dislocation on the electrons can be described 
as a magnetic string located  at the core of the dislocation.
This force is caused by the distorted topology of the real space  
\cite{tk1, io1}. 
In this approximation the crystal lattice is completely lost. 
This model was refined recently \cite{bau1,bau2} by taking into account also 
the distance dependence of the transfer matrix of the tight binding model
in the continuum limit. 
In the context of dislocations the 
interaction energy of two opposite screw dislocations is an 
interesting quantity. This energy was first calculated 
using elasticity theory \cite{na1}. Recently,   
it became possible to carry out ab initio calculations  
on parallel computers 
to calculate the interaction energy of two opposite dislocations for   
special crystals which take also the electron system into 
account (e.g. \cite{ar1}). These calculations 
 neglect the interaction potential between the electrons.
Our analysis will thus also be useful in 
the field of dislocations in solids.   \\

The strategy of the proof that the ground state energy is the same with 
and without the two strings is as follows: 
In section II we show by symmetry considerations that the 
energy spectrum of the two string system 
does not depend on the string separation 
for $d $ strictly positive. This is done by calculating the 
derivate of the energy function with respect to the string separation $d $.
By using the space inversion symmetry of the two string system we 
are able to express this derivate 
as a function which depends on the  
asymptotics of the wavefunction at the two strings. 
This will be shown in subsection A of section II. 
In subsection B  we go further by rewriting this function as 
a differential of the energy with respect to the magnetic 
field and the interaction strength. Thus, we obtain a differential equation 
for the energy function. 
By using the space inversion symmetry 
we obtain from this differential equation 
that the energy spectrum does not depend on the string separation 
$d $. In section III we show that the ground state energy 
is continuous at string separation zero. The technical details of the proof 
are deferred to appendix A. By using the results of section II and section III
we obtain in section IV that the ground state energy and degeneracy 
is the same for the system with or without the two strings. We give a 
conclusion in section V.  
 
\section{The separation distance independence of the energy spectrum}
In order to show the spectral properties mentioned above 
for the infinite system we consider in this paper a high symmetric 
finite system with zero boundary conditions. The boundary 
should have an inversion symmetry. The two strings are arranged in order 
to ensure that the inversion symmetry is indeed conserved. 
In this paper we show the spectral properties 
for this system. We loose the boundary dependence when carrying out 
the limit to an infinite system size. 
  
The Hamiltonian of $ N $ electrons in the background 
of a homogeneous magnetic field $ \vec{B}=-B \vec{e}_z $ ($ B >0 $)
and two magnetic strings of opposite strength $ \pm \phi $ separated by $ d $ 
is given by 
\begin{equation}  
 H_{ss}(\phi,d)  =   \sum^N_{i}\frac{1}{2m}
\bigg(-i \vec{\nabla}_i+\vec{A}(\vec{r}_i) +
  \phi\vec{f}(\vec{r}_i+d\vec{e}_x/2)  
   -\phi\vec{f}(\vec{r}_i-d\vec{e}_x/2 ) \bigg)^2  
 + \frac{\alpha}{2}  \sum_{i\not= j} V^{ee}(\vec{r}_i-\vec{r}_j) \;, \label{10}
\end{equation}
where the convention $ \hbar=1 $, $ c=1$ and $ e=1 $ has been used. 
$ \vec{A}(\vec{r}) $ is the vector potential 
$ \vec{A}=\vec{B} \times \vec{r} /2$.
The function $ \vec{f}(\vec{r}) $ represents  the vector 
potential of a magnetic string with unit flux quantum and is given by  
$ \vec{f}(\vec{r})=\vec{e}_z \times \vec{r}/r^2 $.
 $ V^{ee}(\vec{r}) $ is a potential function. In the following, we 
restrict us at first on 
a homogeneous potential functions of scaling dimension 
$\gamma \in \mathbb{R}$, i.e. 
$ V^{ee}(a \vec{r})=|a|^{-\gamma} V^{ee}(\vec{r})$. 
In the case 
of the Coloumb interaction we have $ \gamma =1$. $ \alpha $ is 
the coupling constant of the interaction potential. 
In the following derivations we consider only systems 
with string strength $ 0 \le \phi < 1$. By using the fact that 
the Hamiltonian (\ref{10})  
with string strength $ \phi $ is unitary equivalent 
to the Hamiltonian with strength 
$ \phi+z, z\in \mathbb{Z}  $ we immediately obtain a generalization of all 
results deduced in this paper to arbitrary $ \phi $. 
The transformation between both systems is given 
by a phase transformation. 

Next, we have to determine the domain of the Hamiltonian $ H_{ss} $ 
(\ref{10}), which can be done in two ways. 
We restrict ourselves to a short outline due to lack of space. 
A much more detailed discussion can be found in the literature cited below.

First, one can determine the self-adjoint domains of 
$ H_{ss} $ by a mathematical extension formalism \cite{ph1}. There are  
many different self-adjoint extensions of $ H_{ss} $. Which of these 
extensions has to be used, depends on the particular  
physical system \cite{aud1}. 
For example, an additional 
 point force such as a Zeeman term at the origin of the string 
can change the extension that has to be used. 

In this paper, we pursue a more physical way of determining the 
domain of (\ref{10}). It consists of  
a regularization of  the magnetic strings to a tube of finite width. 
This yields  
a Hamiltonian with finite regular potentials for which the self-adjoint 
domain of the Hamiltonian is well known. After determining the 
eigenvalues and eigenfunctions of this regularized Hamiltonian, one can  
carry out the limit of tube width zero. This was done first for the 
one string system by Aharonov and Bohm \cite{ah1} and for 
much more general regularizations by Hagen  
in Ref. \onlinecite{ha1}. Similar considerations for the two string 
system were outlined by the author \cite{di2} and can be 
extended easily to the 
many particle (interacting) case. 
We obtain that the physical domain $ {\cal D}[H_{\phi,d}] $ 
of the Hamiltonian $ H_{ss}$ consists of wavefunctions with nodes at the 
positions of the two strings  
\begin{equation}
{\cal D}[H_{\phi,d}]:=\{\Psi(\vec{r})\;|\;  \Psi(\vec{r}) \;
\mbox{is square integrable and} \; \Psi(-d \vec{e}_x/2)=\Psi(d \vec{e}_x/2)=0
\} \,. \label{12}
\end{equation} 
It is then easily seen by carrying out a partial
integration that the Hamilton operator (\ref{10}) is symmetric on this 
domain.  

Next,  we will determine the asymptotics of the eigenfunctions of 
$ H_{ss} $ at the 
position of the two strings. Without loss of generality we determine 
the asymptotics for $ \vec{r}_1 \to \mp d \vec{e}_x/2 $. 
We now seek a solution of the Schr\"odinger equation (\ref{10}) 
of the asymptotic form
\begin{equation}
 \Psi(r)
=\frac{1}{\sqrt{2 \pi}} 
f^{\pm}_p(|\vec{r}_1^{\pm}|  ; \vec{r}_2,..,\vec{r}_N) 
e^{i p \varphi_{\pm}}  \label{15}
\end{equation}
where  $\vec{r}_1^\pm:=
\vec{r}_1 \mp d \vec{e}_x/2$ using cylindrical polar coordinates. 
$ f^{\pm}_p $ is the asymptotic wave function 
for $ \vec{r}_1 \to \mp d \vec{e}_x/2$. 
Then one gets for the eigenvalue equation
in polar coordinates at $ \vec{r}_1 \approx \mp d \vec{e}_x/2$ 
\begin{align}
&  -\frac{1}{2m}\left(\partial^2_{|\vec{r}_1^\pm | }+
\frac{1}{|\vec{r}_1^\pm|}\partial_{|\vec{r}_1^\pm|} -\frac{(p \mp \phi)^2}{|\vec{r}_1^\pm|^2}  \right) 
f^\pm_p(|\vec{r}_1^\pm|;\vec{r}_2,..,\vec{r}_N )   \nonumber \\
& +\sum_{p'} 
\bigg(M_{p,p'}(|\vec{r}_1^\pm|)f_{p'}(|\vec{r}_1^\pm|)
+ 
D_{p,p'}(|\vec{r}_1^\pm|)\partial_{|\vec{r}_1^\pm|} 
\bigg)   
f^\pm_{p'}(|\vec{r}_1^\pm|;\vec{r}_2,..,\vec{r}_N )
= 
E \, f^\pm_{p}(|\vec{r}_1^\pm|;\vec{r}_2,..,\vec{r}_N ) \,. \label{18}
\end{align}  
$ M_{p,p'} $ is a multiplication  operator 
which scales as  $ O(1/|\vec{r}_1^\pm|)$ for 
$ |\vec{r}_1^\pm| \to 0 $ or a differential operator with 
respect to the coordinates $ \vec{r}_2,..,\vec{r}_N $ . 
Similarly $ D_{p,p'} $ is a 
multiplication operator which scales as 
$ O(|\vec{r}_1^\pm|^0) $. Both of these operators reduce 
the scaling power of 
the function $ f^\pm_{p} $ at most by one.   
The asymptotic eigenvalue equation (\ref{15}) and the restriction 
on the wavefunction  (\ref{12}) then immediately leads to 
\begin{equation}
\label{20}
 f^\pm_{p}= v^\pm_{p}(\vec{r}_2,..,\vec{r}_N)\left( 
\delta_{p,0}|\vec{r}_1^\pm|^{\phi}
+\delta_{p,\pm 1}|\vec{r}_1^\pm|^{1-\phi}\right)+O(|\vec{r}_1^\pm|)\,.
\end{equation} 
for $ d >0 $. 
Here $ v^\pm_{p} $ is a finite function of 
$ \vec{r}_2,.., \vec{r}_N  \not=\pm  \vec{d}/2 $. We see from (\ref{20}) that 
the eigenfunctions have not to be continuous in general at $ d=0 $ for 
$ 0< \phi < 1 $. The same holds also for the eigenvalues. 
  
In the following, we  denote by $ \Psi^n_{\phi,d} $ the normalized
eigenstates with energies  $ E^n(\phi,d)$ labelled by $ n $. 
With the help of the current operator 
$ 
\hat{\vec{J}}_{\phi,d}(\vec{r}) =  \sum_i
\delta(\vec{r}-\vec{r}_i) (-i \vec{\nabla}_i+\vec{A}(\vec{r}_i) 
+ \phi\vec{f}(\vec{r}_i^-)
  -\phi\vec{f}(\vec{r}_i^+)
)/m $ 
 (the physical current being  the real part of the
 expectation value of this current operator) we get 
\begin{equation}
 \partial_{\xi} E^{n}=  
\int \nts{1} d^2r (\partial_{\xi} \vec{F}_{\phi,d}(\vec{r})) 
\langle \hat{\vec{J}}(\vec{r}) \rangle^n_{\phi,d} +S^{n}_{\xi}            \label{30}
\end{equation}
with
\begin{equation}
 \vec{F}_{\phi,d}(\vec{r})=   
\frac{\phi}{m} \left(\vec{f}(\vec{r}^-)-\vec{f}(\vec{r}^+) \right) 
\label{40}
\end{equation}
and 
\begin{equation} 
S^{n}_{\xi}(\phi)= \sum\limits_{\sigma \in \{\pm\}} \sum_{i=1}^N 
\int d^2r_1.. d^2r_{i-1}d^2r_{i+1}.. 
d^2r_{N}  
     \left[\,\int\limits_{S_{\sigma \vec{d}/2}} \nts{3} 
 d\vec{o}_i \left((\vec{\nabla}_i \Psi^n_{\phi,d})^* 
\partial_{\xi}\Psi^n_{\phi,d}-(\Psi^n_{\phi,d})^* \vec{\nabla}_i
\partial_{\xi}\Psi^n_{\phi,d}\right)\right]   \label{45}  
\end{equation}
where $ \xi $ equals $ d $ or $ \phi $, respectively.
Here, we denoted  $ \langle \Psi^n_{\phi,d}|
 \hat{\vec{J}}_{\phi,d}(\vec{r})
|\Psi^n_{\phi,d}\rangle $ by
$ \langle \hat{\vec{J}}(\vec{r}) \rangle^n_{\phi,d} $. 
The integration contour 
of the surface integral  
$ \int_{S_{\sigma \vec{d}/2}} d\vec{o}_i $ is  
an infinitesimally small circle around the 
string position $ \vec{r}_i=\sigma  d\vec{e}_x/2$. 
With the help of the asymptotics (\ref{20}) of the eigenfunctions, 
we can show easily that $ S^{n}_\phi $ is zero. This is not the case for 
$ S^{n}_d $. The reason is that the  
domain (\ref{12}) of the Hamiltonian (\ref{10}) changes for 
different string positions, i.e. the derivation of the 
eigenfunction $ \Psi^n_{\phi,d} $ with respect to $ d $ 
can be divergent at the string position and thus is not in the  
domain of  the Hamiltonian (\ref{10}). This results in  
surface terms $ S^{n}_{d}$. 
With the help of the definition of the restricted overlap function 
\begin{equation} 
O^{n}(\phi)=\sum_{\sigma \in {\pm}}\sigma  \int d^2r_2\dots d^2r_{N} 
(v_0^{\sigma,n}(\vec{r}_2,.,\vec{r}_N))^* 
 v_{\sigma 1}^{\sigma,n}(\vec{r}_2,.,\vec{r}_N)  \label{50}
\end{equation}
we get for $ S^{n}_d $ by using (\ref{10}) 
\begin{equation}
 S^{n}_d(\phi) =   
 N \frac{\pi}{m} \phi
\left(\left(\frac{\phi}{2}-1\right)
O^{n}(\phi)+\left(\frac{\phi-1}{2}\right) (O^{n}(\phi))^*\right)\,.    \label{60}
 \end{equation}
This is a term which is in general not zero.

One can interpret formula (\ref{30}) as the energy variation 
of the electron system caused by an induced electric field 
originating from a variation of the magnetic string field.
The first term in (\ref{30}) originates from the force of the induced 
electric field on the electrons outside the strings and the last 
term from the force inside the magnetic strings. 
This is an astonishing result because from (\ref{20}) we see that 
the electron density at the positions of the magnetic strings is zero. 
Nevertheless, one can have a finite current density at the positions of the 
strings resulting in this additional term. 
We should mention that this additional term was erroneously omitted 
in \cite{di2}. 

\subsection{Space inversion symmetry of the system} 
In the following, we will show by using the space inversion symmetry
of the two string system that the energy spectrum 
does not depend on the string distance $ d $ for $ d >0 $. To achieve  
this result we will first express  $ \partial_{d} E^n(\phi,d) $ 
in (\ref{30}) as a pure surface term. Due to the space inversion symmetry, 
we have the following relation for the eigenfunctions of $ H_{ss} $ 

\begin{align}
&\Psi_{1-\phi,d}^n(\vec{r}_1,.., \vec{r}_N)  \label{70}\\
& 
= \Psi^n_{\phi,d}(-\vec{r}_1,..,-\vec{r}_N)
e^{-i \sum_{i} \nts{2} \mbox{ \scriptsize arg}[\vec{r}_i^-]}
e^{+i \sum_{i}\nts{2} \mbox{ \scriptsize arg}[\vec{r}_i^+]} \,.
\nonumber 
\end{align}
Here $\mbox{ arg}[\vec{r}] $ is the angle between $ \vec{r} $ 
and the $ x $ axis.
The wavefunction $ \Psi_{1-\phi,d}^n(\vec{r}_1,.., \vec{r}_N)$ is 
an eigenfunction of $ H_{ss} $ with string strength $ 1-\phi $, i.e. 
\begin{equation} 
E^n(\phi,d)=E^n(1-\phi,d)\,.              \label{75}
\end{equation} 
Thus we denote the related states (\ref{70}) by the
same label $ n $.
The current expectation value is given by  
\begin{equation}
\langle \hat{\vec{J}}(\vec{r}) \rangle^n_{1-\phi,d}=
-\langle \hat{\vec{J}}(-\vec{r}) \rangle^n_{\phi,d}\,.   \label{80}
\end{equation}
In summary, due to the different signs of the strengths of the two strings 
we obtain an inversion symmetry of the energy spectrum 
with respect to the strength $ \phi=0.5 $. 

By using the symmetry of the wave functions (\ref{70}) in order to get 
the corresponding symmetries for the values of the wavefunctions 
at the strings (\ref{20}) we obtain for $ S^{n}_d(1-\phi) $ by using 
(\ref{50}) and (\ref{60})   
\begin{equation}
 S^{n}_d(1-\phi) =    
 N \frac{\pi}{m} \left(\phi-1\right)
\left(\frac{\phi}{2} \, 
O^{n}(\phi)+\left(\frac{\phi+1}{2}\right) 
(O^{n}(\phi))^*\right) \,.   \label{90}
 \end{equation}

In the
following, we calculate the derive of the state energy $ E^n(\phi,d) $ 
with respect to $ d $. By using 
$ \vec{F}_{\phi,d}(\vec{r})=\vec{F}_{\phi,d}(-\vec{r}) $, 
$ E^n(\phi,d)=E^n(1-\phi,d) $ and (\ref{30}), (\ref{80}) 
we get
\begin{eqnarray} 
   \partial_{d} E^n(\phi,d)    
& = &       -\frac{\phi}{1-\phi} \;
(\partial_{d} \vec{F}_{\phi,d}(\vec{r})) 
\langle \hat{\vec{J}}(\vec{r}) \rangle^n_{1-\phi,d}+S^{n}_{d}(\phi) 
 \nonumber \\
&  = &  -\frac{\phi}{1-\phi} \, \partial_{d}  E^n(\phi,d) 
+\frac{\phi}{1-\phi} S^{n}_{d}(1-\phi) +S^{n}_{d}(\phi)\,,   \label{100} 
\end{eqnarray} 
for $ d>0 $. 
With  the definition of the overlap function $ O^n(\phi) $ 
(\ref{50}) and by using (\ref{60}), (\ref{90}) we get 
\begin{equation}
\frac{\phi}{1-\phi} S^{n}_{d}(1-\phi) +S^{n}_{d}(\phi)=
 -N \frac{2\pi}{m} \phi\, \mbox{Re} \, [O^n(\phi)]  \,.         \label{110}
\end{equation}
Now we insert (\ref{110}) in (\ref{100}) to get 
\begin{equation} 
 \partial_{d} E^n(\phi,d)=
 N \frac{2\pi}{m} \phi\,(\phi-1) \mbox{Re} \, [O^n(\phi)]  \,.       
  \label{115}
\end{equation}
Thus, we obtain a formula which relates 
the distance dependence of the energy spectrum to the asymptotic values 
of the wavefunction at the two magnetic strings. 
When neglecting the surface term on the right hand side 
which stems from the domain subtleties of the  
Hamilton operator $ H_{ss} $ mentioned above  we obtain the invariance of 
the energy spectrum as a function of the string distance $ d $. 
In the following, we will show that this term is in fact zero. 
This will be done by deriving a virial theorem for the two string system.

\subsection{Generalized virial theorem for the two string system}

In order to derive this virial theorem we consider the operator 
$ \vec{\Pi}_i=(-i \vec{\nabla}_i+\vec{A}+
 \phi\vec{f}(\vec{r}^-_i)- \phi\vec{f}(\vec{r}^+_i))  $ of the $i$th electron.  
This operator fulfills the following commutation relation 
\begin{equation}
[\Pi_{x_i},\Pi_{y_i}]= i (B- \phi\, \delta(\vec{r}^-_i)
+\phi\, \delta(\vec{r}^+_i)) \,.  \label{120} 
\end{equation}  
For $ N=1$, the ladder operators 
$ \Pi^+_i=(\Pi_{x,i}+i\Pi_{y,i})/2m $  
($ \Pi^-_i=(\Pi_{x_i}-i\Pi_{y_i})/2m  $), when acting on  an 
eigenfunction of $ H_{ss} $, raise (lower) the 
eigenvalue (remember that the eigenfunctions of $ H_{ss}$ are zero at the 
strings for $ \phi \not=0 $). This is well known for a single electron  
in a homogeneous magnetic field without strings.
However, a generalization of this procedure  
in the presence of strings is rather difficult. 
This is due to the fact that 
the wavefunctions created by acting with the ladder 
operators $ \Pi^{+}_i$ and $\Pi^{-}_i$ on an 
eigenfunction need not be in the domain ${\cal D}[H_{\phi,d}] $ (\ref{12}) 
of $ H_{ss} $ for $ \phi \not=0 $. 
This can be shown easily for the known ground state wavefunctions 
of the one and two string systems \cite{di2}.

In order to express the surface term on the right hand side of 
equation (\ref{115}) as a function of the energy $ E^n(\phi,d) $ 
we consider the operator $ \sum_{i=1}^N \{\vec{r}_i, \vec{\Pi}_i\} $ 
where $ \{\cdot,\cdot\} $ is the anti-commutator. 
Then, by using (\ref{120}) and the homogeneity of the interaction function 
$ V^{ee} $ we obtain  
\begin{align}
&i \left[H_{ss},\sum_{i=1}^N \left\{\vec{r}_i, \vec{\Pi}_i\right\}\right] =
4 H_{ss} +  (\gamma-2) \alpha  
\sum\limits_{i \not= j} V^{ee}(\vec{r}_i -\vec{r}_j) \nonumber  \\
& +\frac{1}{m}\Bigg[\sum_{i=1}^{N} B \left[\vec{r}_i 
{\begin{array} {c}  \vspace{-0.4cm} \times \vspace{0.05cm} \\  ,
\end{array}}
 \vec{\Pi}_i\right] 
+\left[\frac{\vec{r}_i}{2}  {\begin{array} {c}  
\vspace{-0.4cm} \times \vspace{0.08cm} \\  ,\end{array}} 
\left\{\phi \,\delta(\vec{r}_i^+)- \phi \,\delta(\vec{r}_i^-), \vec{\Pi}_i
\right\}\right]\Bigg]\,. \label{125}
\end{align} 
Here $ \vec{A} \times \vec{B}-\vec{B}\times \vec{A} $ is denoted by 
$ [\vec{A} {\begin{array} {c}  
\vspace{-0.4cm} \times \vspace{0.08cm} \\  ,\end{array}} \vec{B}] $ 
for two vector operators $ \vec{A} $, $\vec{B}$.  

In the following, we will calculate the expectation values 
of the various operators in (\ref{125})  with respect to the eigenfunction 
$ \Psi^n_{\phi,d} $. The expectation values of the two operators 
which contain only surface terms are given by 
\begin{eqnarray}
 i 
\big\langle\big[H_{ss},\sum_{i=1}^N \left\{\vec{r}_i, \vec{\Pi}_i\right\}\big]
\big\rangle^n_{\phi,d} & = &  
 N \frac{4\pi d}{m} \phi (1-\phi)\mbox{Re}[O^n(\phi)] \,,  \label{130}  \\ 
 \frac{1}{2m} \sum\limits_{i=1}^{N}
\left\langle\left[\vec{r}_i  {\begin{array} {c}  
\vspace{-0.4cm} \times \vspace{0.08cm} \\  ,\end{array}} 
\left\{\phi \,\delta(\vec{r}_i^+)- \phi \,\delta(\vec{r}_i^-
), \vec{\Pi}_i
\right\}\right]\right\rangle^n_{\phi,d}    
 & = &  N \frac{2\pi d}{m} \phi (1-2\phi)\mbox{Re}[O^n(\phi)] \,. \label{135}
\end{eqnarray} 
Now we have everything that is required for the derivation of the energy 
formula. With the help of (\ref{10}), (\ref{115}), (\ref{130}) and 
(\ref{135}) we obtain the following differential equation for the energy 
function $ E^n(\phi,d) $
\begin{equation}
\frac{d}{(\phi-1)} \frac{\partial}{\partial d} E^n(\phi,d) =
4 \left( 1 - B \frac{\partial}{\partial B}+ \frac{(\gamma-2)\alpha}{2}
\frac{\partial}{\partial \alpha } \right)E^n(\phi,d)  \label{140}
\end{equation}    
This is a remarkable result stating that the energy is 
a homogeneous function of the possibly rescaled variables 
$ \alpha $, $ B$ and $d$. 
We now come to the conclusion that the energy function does not depend 
on the string distance $ d $ for $ d>0 $. This can be seen 
by the use of the relation (\ref{75}). From this mirror symmetry
relation of the spectrum we get that $ E^n(\phi,d) $ fulfills 
also the differential equation (\ref{140}) in the case of the substitution 
$ \phi \rightarrow (1-\phi) $ in the left hand side of (\ref{140}). 
Thus we obtain 
$ (\partial/\partial d) E^n(\phi,d)/(\phi-1)=
-(\partial/\partial d) E^n(\phi,d)/\phi $. This equation can only be 
fulfilled  when $ (\partial/\partial d) E^n(\phi,d)=0 $ identically. 
Summarizing, we obtain: 
\begin{align}
& \frac{\partial}{\partial d} \, E^n  =  0  \;,         \label{150} \\
&  E^n  - B \frac{\partial}{\partial B}E^n+ 
\frac{(\gamma-2)\,\alpha}{2}\; 
\frac{\partial}{\partial \alpha }E^n   =  0  \,.   
\label{160}
 \end{align} 
This is the main result of this section. It is valid only in the case 
of a homogeneous interaction potential $ V^{ee} $. 

The first equation states that 
the energy spectrum does not depend on the string separation distance  
for $d>0$. 
We get the constraint $ d >0 $ because 
we used the asymptotic formula (\ref{20}) for the 
eigenfunctions in the derivation above.
This formula is only valid for a finite positive string distance  
because the eigenfunctions and eigenvalues 
are not continuous in general at $ d=0 $. Nevertheless, we will show 
in the following section that the ground state energy 
as well as the ground state degeneracy are 
continuous as a function of $ d $ at $ d=0 $. 
 
The differential equation (\ref{160}) states that $ E^n(\phi,d) $ 
is a homogeneous function of the magnetic field $ B $ and the 
interaction coupling constant $ \alpha^{(2-\gamma)/2} $. 
For $ \phi=0 $ and $ B=0 $ this equation is the well known 
quantum mechanical version of the virial theorem \cite{fi1}.

The above results for the homogeneous two particle 
interaction potential $ V^{ee} $ can be 
generalized. It is easily seen 
from the derivation above that (\ref{150}) and (\ref{160}) are 
also valid for homogeneous many particle potentials 
$ \alpha V(\vec{r}_1,..,\vec{r}_N) $. 
Furthermore, (\ref{150}) and (\ref{160}) can be generalized 
to the more general 
case of an inversion symmetrical potential, i.e. 
$ \alpha V(\vec{r}_1,\dots,\vec{r}_N)=\alpha \,V(-\vec{r}_1,\dots,-\vec{r}_N) 
$ by substituting 
$ (\gamma/2-1) 
\alpha (\partial/\partial \alpha) E^n \rightarrow 
-\alpha \langle  
(V(\vec{r}_1,..,\vec{r}_N)+
\sum_{i=1}^N (\vec{r}_i \cdot \vec{\nabla}_i) V(\vec{r}_1,..,\vec{r}_N)/2)  \rangle^n_{\phi,d} $ in the 
last term on the right hand side of  (\ref{160}). 
We should point out that the physically relevant 
interaction potential of the jellium system 
is exactly of that form. 
 
\section{Ground state energy for small string separation}
In this section we will show the continuity of the ground state energy for 
small string separations.  
In the following we briefly outline the strategy of the proof deferring 
the details to appendix A.

Given a (degenerate) ground state wavefunction 
of the electron system without strings (a ground state wavefunction of 
$ H_{ss}(0,0) $), we construct wavefunctions which 
depend on the string separation $ d $.  These wavefunctions  have  
expectation values with respect to the Hamiltonian $ H_{ss}(\phi,d) $  
which converge towards the ground state energy of $ H_{ss}(0,0) $ for 
$ d \to 0 $. 
By using the Rayleigh-Ritz variational principle we obtain 
$  \lim_{d \to 0} (E^{\mbox{\tiny GS}}_{\phi,d}
-E^{\mbox{\tiny GS}}_{0,0}) \le 0 $ (\ref{570})
where $ E^{\mbox{\tiny GS}}_{\phi,d} $ is the ground state energy 
of $ H_{ss} $.
A rigorous proof of the continuity of the ground state energy for 
small string separations, requires the proof of the reverse, too. Given 
a ground state wave function of $ H_{ss}(\phi,d) $ we will construct 
wavefunctions (depending on string distance $ d $) for which we 
calculate the expectation 
value with respect to the Hamiltonian $ H_{ss}(0,0) $. 
We will show that the  difference of this expectation value 
and the ground state energy of $ H_{ss}(\phi,d) $  converges 
to zero for $ d \to 0 $. By using the 
Rayleigh-Ritz variational principle we obtain 
$  \lim_{d \to 0} (E^{\mbox{\tiny GS}}_{\phi,d}
-E^{\mbox{\tiny GS}}_{0,0}) \ge 0 $ (\ref{a40}). 

By combining (\ref{570}) and (\ref{a40})  
we obtain that the ground state 
energy of $ H_{ss}(\phi,d) $ is continuous at $ d=0 $, i.e. 
\begin{equation}
\lim_{d \to 0} (E^{\mbox{\tiny GS}}_{\phi,d}
-E^{\mbox{\tiny GS}}_{0,0}) =  0   \;.  \label{1000}
\end{equation}
From the explicit construction carried out in appendix A 
one further obtains the continuity of the ground state degeneracy at $d=0$.  

\section{Ground state energy without restrictions on the string 
separation}
In section II, we have shown that the energy spectrum does not depend on the 
string separation. In section III, the continuity of the ground state energy 
at $ d=0 $ has been shown.  
Using Eq. (\ref{150}) and Eq. (\ref{1000}) 
we obtain that the ground state energy is the same for the system 
with or without the two strings, i.e. 
\begin{equation}
E^{\mbox{\tiny GS}}_{\phi,d}=E^{\mbox{\tiny GS}}_{0,0} \,,  \label{600} 
\end{equation}
Furthermore, we obtain from the remark below  (\ref{1000}) that the 
ground state degeneracy is also the same.

We should mention that this result is to our understanding by no means  
trivial which can be seen by considering 
the ground state energy of an interacting electron system 
in the background of one string. Let us suppose that 
the ground state degeneracy is lifted by the 
interaction between the electrons for the system 
without the string. This ground state 
wavefunction is changed when adiabatically switching on the magnetic string.
During this process the angular momentum of the wavefunction in $z$-direction 
remains constant. When the string reaches one flux quantum the ground state 
of this system is connected with the ground state of the system without 
the string by a phase transformation. Thus, the angular momentum of 
the two ground states for the system with zero flux quantum and one flux quantum differs by $ N \hbar$. This means that the ground states 
of the system with zero string flux and with string flux one are not connected 
in the energy spectrum as a function of the string strength. 
Then we obtain a crossing of the energy values of these two states 
at values between zero and one. 
This crossing should in general happen at an energy above the ground 
state energy of the system without the magnetic string. 
Thus, we obtain that the ground state 
energy is in general not the same with or without the string. 
The main difference between the one string system and the system with two 
opposite strings considered in this paper is that the net flux is 
zero for the two string system. Thus, one could suppose that the results 
above are true in general for a many string system with 
resulting net zero flux. This is an open question which can  not be 
answered by the symmetry considerations carried out in this paper.   
 
At last we compare the results derived in this paper with 
the corresponding results for the non-interacting 
($N=1$) two string system in a homogeneous magnetic field 
considered in  Ref. \onlinecite{di2}. There we obtain 
that only the energy of the wavefunction which is constant on both strings 
increases when switching on the magnetic strings. 
The energies of the rest of the ground states are constant. 
Due to the infinite degeneracy of the ground state we obtain 
the correspondence between the results in this paper and in Ref. 
\onlinecite{di2} concerning the ground state energy and the ground state 
degeneracy of the two string system.  
The increase of the energy of one of the ground states 
when switching on the strings is a pure edge 
effect which is linked to the property that 
the degeneracy of the non-interacting system is proportional 
to its area.

\section{Conclusion}
We discuss in this paper the problem of the ground state energy 
calculation for an infinitely extended  system of  
interacting electrons in a homogeneous magnetic field 
and in the  background of two magnetic strings with opposite flux.
We treat at first an interaction potential for the electrons which 
is homogeneous. 
By using the 
inversion symmetry of the two string system and further some commutator 
algebra  we  
obtain two equations. 
The first equation (\ref{150}) states that the energy spectrum does 
not depend on the separation distance of the two strings for strictly positive 
string separation. 
The second equation (\ref{160}) has the same shape as the 
virial theorem equation for the system without 
strings. It contains only differentials 
with respect to the magnetic field and the interaction strength.
This equation gives evidence on the homogeneity 
of the energy function as a function of the 
rescaled magnetic field and interaction strength.     
Next, we showed by using the Rayleigh Ritz variational principle 
that the ground state energy and the ground state 
degeneracy are continuous functions at zero string separation. 

By using these properties, we obtain the main result of this paper: 
The ground state energy of the interacting system 
is the same with and without the two magnetic strings.  
This is an answer 
to a question that arose in the context of determining  
a theory for the fractional quantum Hall effect with a finite Green's 
function  \cite{di2} but will also be useful in other fields of physics, 
as for example in the dislocation theory of solids. 

We should point out that the results concerning the 
independence of the energy spectrum with respect to the string separation 
and the invariance of the ground state energy are also valid 
for the more general class of inversion symmetrical 
many particle interaction potentials for indistinguishable particles.
For example, the two particle interaction potentials 
which depend on the distance between the electrons are of that form.  
Furthermore, we obtain the validity of the results also for the jellium model.
We can generalize our findings to the case of bosons. 

\bigskip

We would like to thank Christian Joas, M. Semmelhack, 
and Sven Gnutzmann for helpful 
discussions during the course of this work. This work has been supported 
by the DFG-Schwerpunktprogramm Quanten-Hall-Systeme.  

\begin{appendix}

\section{Proof of the continuity of the ground state energy for small 
string separation}

In this appendix, we give a formal proof that the  
the ground state energy and the ground state degeneracy are  continuous 
functions of $ d $ for $ d \to 0 $ following the strategy 
outlined in section III .

\subsection{Proof of 
\boldmath $ \lim_{d \to 0} (E^{\mbox{\tiny GS}}_{\phi,d}
-E^{\mbox{\tiny GS}}_{0,0}) \le  0 $ \boldmath} 

We first construct approximate  ground state 
wavefunctions of $ H_{ss}(\phi,d) $ from the ground state of 
$ H_{ss}(0,0) $.
To this end, we define the following auxiliary function
\begin{equation}        \label{490} 
P_{\phi_d,d}:=       
\begin{cases}
(r^-)^{\phi_d} & \text{for} \quad x\le -d/2 \,,\\
\frac{y^2}{(|y|+|x^-|)^{2-\phi_d}} & \text{for} \quad -d/2\le x<0 \,,\\
\frac{y^2}{(|y|+|x^+|)^{2-\phi_d}} & \text{for} \quad 0\le  x<+d/2 \,,\\
(r^+)^{\phi_d} & \text{for} \quad d/2 \le x \,. 
\end{cases}
\end{equation}
which is continuous on $\mathbb{R}^2 $.
Let $ \Psi^{\mbox{\tiny GS}}_{0,0} $ be an arbitrary ground state 
wavefunction of $ H_{ss}(0,0) $.
The corresponding approximate ground
state of $ H_{ss}(0,0) $ is given by 
\begin{equation}\label{500} 
 \Psi_{\phi,d}:=\frac{1}{N(\phi,d)} \left(\prod_{i=1}^N e^{i \phi 
\mbox{\scriptsize arg}[\vec{r}_i^+]} 
 e^{-i \phi \mbox{\scriptsize  arg}[\vec{r}_i^-]}
P_{\phi_d,d}(\vec{r}_i)\right)  \Psi^{\mbox{\tiny GS}}_{0,0} 
\end{equation}
with $ N(\phi,d) $ a normalization factor defined by 
$ \int d^2r |\Psi_{\phi,d}(\vec{r})|^2=1 $ and
\begin{equation}
\phi_d=\frac{1}{\sqrt{|\ln(d)|}} \;. \label{505}   
\end{equation}  
Due to the non-differentiability of $ \Psi_{\phi,d} $ on the axes $ x=-d/2 $, 
$ x=0 $ and $ x=d/2 $ 
we get that $ \Psi_{\phi,d} $ is not in the domain of 
$ H_{ss}(\phi,d) $. Nevertheless,  $\Psi_{\phi,d} $ 
is continuous on  $\mathbb{R}^2 $. It can be shown that  
the wavefunction $\Psi_{\phi,d} $ can be smoothed in a way that 
its expectation value with  respect to $ H_{ss}(\phi,d) $ remains 
unchanged up to an infinitesimal amount and lies in the domain of 
$ H_{ss}(\phi,d) $. 
By definition, it is immediately clear that 
$ \lim_{d \to 0} N(\phi,d)=1 $. In the following, we show that 
$ \lim_{d \to 0}  \langle \Psi_{\phi,d} | H_{ss}(\phi,d) | 
\Psi_{\phi,d} \rangle 
= E^{\mbox{\tiny GS}}_{0,0} $. We restrict 
the discussion to the one particle case in order to simplify the notation. 
The following results can be generalized easily to the many particle case. 
With the help of 
\begin{eqnarray}
T_1 & := &\frac{1}{N(\phi,d)}\frac{1}{2m}\langle   
\Psi^{\mbox{\tiny GS}}_{0,0} |
 (\vec{\nabla} P_{\phi_d,d}(\vec{r})) 
 (\vec{\nabla}  P_{\phi_d,d}(\vec{r}))|
\Psi^{\mbox{\tiny GS}}_{0,0} \rangle \,, \label{510} \\
T_2 & := & -\frac{1}{N(\phi,d)}\frac{1}{2m}\langle  
\Psi^{\mbox{\tiny GS}}_{0,0} | P_{\phi_d,d}(\vec{r}) 
(\vec{\nabla} P_{\phi_d,d}(\vec{r}))
\vec{\nabla}+\vec{\nabla} P_{\phi_d,d}(\vec{r})  
(\vec{\nabla} P_{\phi_d,d}(\vec{r}))
|\Psi^{\mbox{\tiny GS}}_{0,0} \rangle \,, 
   \label{520}   \\
T_3 & := & \frac{1}{N(\phi,d)}\frac{1}{i} 
\frac{1}{m}\langle 
\Psi^{\mbox{\tiny GS}}_{0,0} | 
P_{\phi_d,d}(\vec{r}) 
(\vec{A}(\vec{r})               
(\vec{\nabla}
P_{\phi_d,d}(\vec{r}))|
\Psi^{\mbox{\tiny GS}}_{0,0} \rangle   \label{530}
\end{eqnarray}
we obtain $ \langle \Psi_{\phi,d} | H_{ss}(\phi,d) | 
\Psi_{\phi,d} \rangle=E^{\mbox{\tiny GS}}_{0,0}+T_1+T_2+T_3 $. 
Thus, we have to show that $ \lim_{d \to 0 } (T_1+T_2+T_3) =0 $. 

The expectation values in (\ref{510})-(\ref{530}) stand for a two dimensional  
integration over the plane. In the following, we will show  
that $ \lim_{d \to 0 } T_1=0 $ is fulfilled 
separately for the four  different integration areas $ x<-d/2 $,  
$ -d/2<x<0$, $ 0<x<d/2 $ and $ d/2<x $. It 
is immediately clear 
from the definitions (\ref{490})-(\ref{530})
that $ \lim_{d \to 0 } T_1=0 $  is fulfilled for the integration 
area $ x<-d/2$ and 
$ x>d/2$. Now we will show first  
that $ \lim_{d \to 0} T_1=0  $ is fulfilled for the integration 
area  $ -d/2 <x<0 $ and 
$ y>0 $. 
In this area, we get 
\begin{equation} 
(\vec{\nabla} P_{\phi_d,d})^2=\frac{
2(2+\phi_d^2-2\phi_d) y^4}{(y+|x^-|)^{6-2\phi_d}}+
\frac{4 \phi_d \, x^- y^3+4(x^-)^2 y^2}{(y+|x^-|)^{6-2\phi_d}} \,.\label{550}
\end{equation}
Since $ \Psi^{\mbox{\tiny GS}}_{0,0} $ is continuous 
differentiable it is sufficient to show 
that $ \lim_{d\to 0} \int_{-d/2<x<0} d^2r 
(\vec{\nabla}  P_{\phi_d,d}(\vec{r}))^2  =0 $. As an example we show the 
asymptotic vanishing  of  the 
first term in (\ref{550}). The vanishing of the last term 
in (\ref{550}) works similarly. By carrying out the integration we get 
for the first term 
\begin{equation}
 \int\limits_{-d/2<x<0,y>0} \nts{8} d^2r \, 
\left[(\vec{\nabla} P_{\phi_d,d})^2\right]_1
= -\frac{(2+\phi_d^2-2\phi_d)}{(5-2\phi_d)\phi_d}
\Big[(y_0+d/2)^{2\phi_d}-(d/2)^{2\phi_d}-
 (y_0)^{2\phi_d} 
-2 \phi_d 
\int\limits_{0}^{y_0}dy\, 
\frac{(y+d/2)^4 -y^4}{(y+d/2)^{5-2\phi_d}}\Big]   \label{560}
\,.  
\end{equation}
Here $  y_0 $ is the y-coordinate of the upper edge of the 
sample (without loss of generality we assume a square sample 
with center at the origin).     
By a binomial expansion of the first term in the square brackets 
on the right hand side of (\ref{560}) and the definition of 
$ \phi_d $ in (\ref{505}) we find that the 
sum of the first three terms on  the right hand side is zero for 
$ d\to 0 $. The last term in the square brackets in (\ref{560}) 
converges by a factor 
$ \phi_d $ faster to zero than the sum of the first three terms. By carrying 
out a similar analysis for the last term in (\ref{550}) we get 
 $ \lim_{d\to 0} \int_{-d/2<x<0, y>0} d^2r (\vec{\nabla} P_{\phi_d,d})^2=0 $.  
Due to the mirror symmetry 
according to the $y$-axis and $x$-axis of $ P_{\phi_d,d} $ 
we then find also that
\begin{equation}
\lim_{d\to 0} \int\limits_{-d/2<x<d/2} d^2r \; (\vec{\nabla} P_{\phi_d,d})^2=0
\label{565}
\end{equation}
and thus $ \lim_{d \to 0} T_1=0  $ for $ -d/2 <x<d/2 $.
At last, we have to show that $ \lim_{d \to 0 }  (T_2+T_3)$. It is  
easily seen from the definitions (\ref{520}) and (\ref{530}) that  
$ T_2 $ and $ T_3 $ converge much faster to zero 
than $ T_1 $. 
Summarizing, we get $ \lim_{d \to 0 }  (T_1+T_2+T_3)=0 $. 

Now we obtain by the help of  
$ \lim_{d \to 0 } \langle \Psi_{\phi,d} | H_{ss}(\phi,d) | 
\Psi_{\phi,d} \rangle=E^{\mbox{\tiny GS}}_{0,0} $ shown above 
and the Rayleigh-Ritz variational principle that  
\begin{equation}
 \lim_{d \to 0} (E^{\mbox{\tiny GS}}_{\phi,d}
-E^{\mbox{\tiny GS}}_{0,0}) \le 0  \,.     \label{570}
\end{equation}
Furthermore, we obtain from the definition of $ \Psi_{\phi,d} $
(\ref{500})  
that the number of orthogonal states with an energy lower than or equal to 
$ E^{\mbox{\tiny GS}}_{0,0} $ is larger than or equal to 
the degeneracy of the ground state energy $ E^{\mbox{\tiny GS}}_{0,0} $ for 
$d \to 0 $. 

\subsection{The proof of 
\boldmath $ \lim_{d \to 0} (E^{\mbox{\tiny GS}}_{\phi,d}
-E^{\mbox{\tiny GS}}_{0,0}) \ge  0 $ \boldmath} 

Now we have to show the reverse, i.e. given the 
(degenerate) ground state 
wavefunction $ \Psi^{\mbox{\tiny GS}}_{\phi,d} $ of energy 
$ E^{\mbox{\tiny GS}}_{\phi,d} $, we construct an approximate 
ground state  wavefunction 
 $  \Psi_{0,0} $ for $ H_{ss}(0,0) $ 
(\ref{10}). This is done by defining $ \Psi_{0,0} $ as in (\ref{500}) with 
substitutions $ \Psi^{\mbox{\tiny GS}}_{0,0} \rightarrow 
\Psi^{\mbox{\tiny GS}}_{\phi,d} $, 
$ \mbox{arg}[\vec{r}_i^-] \rightarrow -\mbox{arg}[\vec{r}_i^-]$ and 
$ \mbox{arg}[\vec{r}_i^+] \rightarrow -\mbox{arg}[\vec{r}_i^+] $. Then 
we have to show that $ \lim_{d \to 0 } \langle 
\Psi_{0,0} | H_{ss}(0,0) | 
\Psi_{0,0} \rangle-E^{\mbox{\tiny GS}}_{\phi,d} =0 $.
It is clear that the proof works analogously to the proof of (\ref{570}). 

The ground state energy $ E^{\mbox{\tiny GS}}_{\phi,d} $ is given by  
\begin{equation}
E^{\mbox{\tiny GS}}_{\phi,d}=\int d^2r [(-i \vec{\nabla}+\vec{A}(\vec{r}) +
  \phi\vec{f}(\vec{r}^-)  
   -\phi\vec{f}(\vec{r}^+))\Psi^{\mbox{\tiny GS}}_{\phi,d}]^*
\,[(-i \vec{\nabla}+\vec{A}(\vec{r}) +
  \phi\vec{f}(\vec{r}^-)  
   -\phi\vec{f}(\vec{r}^+))\Psi^{\mbox{\tiny GS}}_{\phi,d}]\,.  \label{a10}
\end{equation}    
The energy expectation value  
$ \langle \Psi_{0,0} |H_{ss}(\phi,d) |\Psi_{0,0} \rangle $ is given by 
$ E^{\mbox{\tiny GS}}_{\phi,d} +
\tilde{T}_1+\tilde{T}_2+\tilde{T}_3 $
where $ \tilde{T}_1$, $ \tilde{T}_2 $ is given by 
$ T_1 $, $ T_2 $ (\ref{510}), (\ref{520}) with the substitution 
$ \Psi_{00}^{\mbox{\tiny GS}} \rightarrow \Psi_{\phi,d}^{\mbox{\tiny GS}}$.  
$ \tilde{T}_3$ is given by $ T_3 $ (\ref{530}) with the substitutions
$ \Psi_{00}^{\mbox{\tiny GS}} \rightarrow \Psi_{\phi,d}^{\mbox{\tiny GS}}$ and 
$ \vec{A}(\vec{r}) \rightarrow \vec{A}(\vec{r}) +
  \phi\vec{f}(\vec{r}^-)  
   -\phi\vec{f}(\vec{r}^+) $. 
In order to show the inequality $ \lim_{d \to 0} (E^{\mbox{\tiny GS}}_{\phi,d}
-E^{\mbox{\tiny GS}}_{0,0}) \ge  0 $  we have to show that 
$ \lim_{d \to 0} (\tilde{T}_1+\tilde{T}_2+\tilde{T}_3)=0 $.
This can be shown rather easily by using (\ref{565}), (\ref{a10}), 
and further the well known Schwarz inequality 
$ |\int_A d^2r f^*(\vec{r})g(\vec{r}) |^2 \le 
\int_A d^2r |f(\vec{r})|^2 \cdot \int_A d^2r |g(\vec{r})|^2 $ valid for 
$ \mathbb L^2 $ wavefunctions $ f $ and $g $ in the area $ A $.

Summarizing, we obtain 
\begin{equation}
 \lim_{d \to 0} (E^{\mbox{\tiny GS}}_{\phi,d}
-E^{\mbox{\tiny GS}}_{0,0}) \ge  0 \,.      \label{a40}
\end{equation} 
Furthermore, we obtain
that the number of orthogonal states with an energy lower or equal to 
$ E^{\mbox{\tiny GS}}_{\phi,d} $ is larger than or equal to  
the degeneracy of the ground state energy $ E^{\mbox{\tiny GS}}_{\phi,d} $ 
for $d \to 0 $.

 \end{appendix}


\begin{thebibliography}{10}
\bibitem{note}
In the case of an infinite particle number, corresponding to 
an infinite energy, the equality of the energy with and without 
the two strings means that the difference of these two energies which is 
finite due to the locality of the strings is zero.
\end{thebibliography}
\end{document}